\documentclass[pra,showpacs,superscriptaddress,reprint,floatfix]{revtex4-1}
\usepackage{graphicx}

\usepackage{ulem}
\usepackage[usenames]{color}

\newcommand{\bra}[1]{\langle #1 |}
\newcommand{\ket}[1]{| #1 \rangle}

\begin{document}

\title{Anomalous Bloch oscillations in one dimensional parity-breaking periodic potentials}

\author{Giulio Pettini}
\email{pettini@fi.unifi.it}
\affiliation{Dipartimento di Fisica e Astronomia, Universit\`a di Firenze,
and INFN, 50019 Sesto Fiorentino, Italy}

\author{Michele Modugno}
\affiliation{Department of Theoretical Physics and History of Science, UPV-EHU, 48080 Bilbao, Spain}

\affiliation{IKERBASQUE, Basque Foundation for Science, 48011 Bilbao, Spain}

\begin{abstract}
We investigate the dynamics of a wave packet in a parity-breaking one-dimensional 
periodic potential slowly varied in time and perturbed by a linear
potential. Parity is broken by considering an asymmetric double well per unit cell.
By comparing the prediction of the semiclassical dynamics with the full Schr\"odinger solution, we show that Bloch oscillations  are strongly affected by anomalous velocity corrections related to Berry's phase. 
We characterize how these effects depend on the degree of parity breaking of the potential and on the modulation parameters. We also discuss how to measure the effects of the anomalous velocity in current experiments with non-interacting Bose-Einstein condensates in bichromatic optical lattices, under the effect of gravity. 
\end{abstract}

\date{\today}


\maketitle

\section{Introduction}

The semiclassical equations of motion play a fundamental role in the transport theory for particles in crystal lattices, being them electrons in metals and semiconductors \cite{ashcroft}, or ultra-cold atoms in optical lattices \cite{morsch}. One of the most intriguing predictions, is the  occurrence of Bloch oscillations when an additional constant force is present \cite{bloch}. Despite being hardly detectable for electrons in natural crystal lattices as predicted in \cite{bloch}, Bloch oscillations have been observed in a variety of systems, like electronic semiconductor superlattices \cite{superlattices}, optical waveguide arrays \cite{waveguides}, and ultracold atoms in optical lattices \cite{salomon,fermions,bosons}. 
In the latter case, long-lived Bloch oscillations have been observed both
 with thermal clouds \cite{salomon} and with non-interacting degenerate fermions \cite{fermions} and bosons \cite{bosons} loaded in 
sinusoidal periodic potentials under the effect of gravity, whose behavior is well described by  the original semiclassical prediction \cite{bloch}.

However, quantum mechanics has implications leading beyond the standard semiclassical approximation  \cite{ashcroft,bloch}. In particular, for cyclic variations of the parameters, as is the case for Bloch oscillations in quasi-momentum space,  phase dependent effects can be present, as shown in the seminal work by M. Berry \cite{berry}.
The importance of Berry's phase in the wave packet dynamics in periodic potentials was recognized by Chang and Niu \cite{niu},  who derived a set of corrections to the semiclassical equations of motions by means of an effective  Lagrangian approach \cite{karplus}.  The whole set of corrections are now generally called Berry corrections or Berry terms \cite{niu1}. 
As a consequence, anomalous transport may occur in case of parity-breaking potentials, in two and three spatial dimensions. Anomalous effects can be present even in one-dimensional systems, provided that Bloch bands are also subjected to slowly-varying time-dependent perturbations, as 
discussed in \cite{niu1} (see also \cite{yuce,chong,artoni,diener}).

In this paper we investigate the effects of Berry corrections on the Bloch oscillations in one-dimensional (1D) time-dependent parity breaking potentials.
 In particular, we consider the dynamics of a non interacting wave packet in a 1D periodic potential, in 
the presence of 
an additional constant force, as can be obtained for instance with ultracold atoms in an optical lattice subjected to gravity \cite{fermions,bosons}. 
Since in 1D Berry terms may appear only when parity is broken and
in the presence of time-depending bands \cite{niu1,libro}, we consider a potential composed by 
two shifted sinusoidal potentials with commensurate wavelengths, whose amplitudes are modulated in time \cite{parametric}.

We find that parity breaking affects significantly the band structure, already in the static case. Then, in the presence of time-modulations, Bloch oscillations may be dramatically modified 
owing to the presence of the anomalous velocity term in the semiclassical equations. 
By direct comparison with the full solution of the Schr\"odinger equation, we show that
these corrections of the semiclassical equations fully accounts for the center of mass dynamics of the wavepacket. We also characterize how the anomalous effects depend on the degree of parity breaking of the potential and on modulation parameters. Finally, we discuss how the effects of the anomalous velocity can be measured from time-of-flight measurements in current experiments with non-interacting Bose-Einstein condensates in optical lattices, in the presence of gravity.

The paper is organized as follows. In Sect. \ref{sec:anomvel}, we start by recalling the expressions for the semiclassical equations of motions and  the effect of Berry corrections on Bloch oscillations. Then, in Sect. \ref{sec:model}, we discuss the model considered, and the static properties of 
the potential in the presence of parity breaking. The emergence of an anomalous velocity term in the presence of time-dependent energy bands, 
its effect on Bloch oscillation, and the dependence on the system parameters, are discussed in Sect. \ref{sec:bloch}. 
Finally, in Sect. \ref{sec:av} we discuss the relevance of the present results for current experiments with ultracold atoms in optical lattices.
Conclusions and outlook are drawn in Sect. \ref{sec:conclusions}

\section{Anomalous velocity}
\label{sec:anomvel}

Let us consider a quantum particle of mass $m$, represented by a wavepacket $\psi(x)$, in a one dimensional periodic potential $V(x)$ of period $a$, $V(x+a)=V(x)$.
Owing to the Bloch theorem, the system eigenfunctions can be written as $\psi_{nk}(x)=e^{ikx}u_{nk}(x)$, the $u_{nk}(x)$'s having the same periodicity of the potential.
Consequently, the energies are restricted within energy bands
 $\varepsilon_{n}(k)$, that are periodic in the quasimomentum $k$-space with period $2k_{B}=2\pi/a$, $\varepsilon_{n}(k+k_{B})=\varepsilon_{n}(k)$.
 
At the semiclassical level, the particle dynamics can be described in terms of the wavepacket centers $x_{c}$ and $k_{c}$  in coordinate and quasi-momentum space, 
respectively. This holds even when the band structure is modulated in time, provided the variations are adiabatic. In particular, in the presence of 
an additional constant force $F$, and within the one-band approximation (we drop the index $n$), the evolution of the wavepacket
 centers is described by the semiclassical equations \cite{ashcroft,niu}
\begin{eqnarray}
\label{eqmoto1}
&&{\dot{x}}_{c}=\frac{1}{\hbar}\frac{\partial\varepsilon}{\partial k}\Big|_{k=k_{c}}+\frac{\partial A_{k}}{\partial t}
\Big|_{k=k_{c}}-\frac{\partial\chi_{k}}{\partial k}\Big|_{k=k_{c}}\\
&&{\dot{k}}_{c}={F}/{\hbar}
\label{eqmoto2}
\end{eqnarray}
where the Berry vector potential $A_{k}$ and scalar potential $\chi_{k}$ in (\ref{eqmoto1}) are  defined as \cite{niu,niu1}
\begin{equation}
A_{k}(t)=i\frac{2\pi}{a}\langle u_{k}|\frac{\partial}{\partial k}|u_{k}\rangle\;,\quad
\chi_{k}(t)=i\frac{2\pi}{a}\langle u_{k}|\frac{\partial}{\partial t}|u_{k}\rangle,
\label{berry}
\end{equation}
with the prefactor $2\pi/a$ coming from the standard normalization of $u_{k}$'s in the unit cell, $\langle u_{k}|u_{k'}\rangle=(a/2\pi)\delta_{kk'}$, at each time $t$ 
\cite{callaway}. We recall that Zak-Berry phase $\gamma$ is defined as the integral over the first Brillouin zone of the vector potential $A_{k}$ \cite{libro}
\begin{equation}
 \gamma=i\frac{2\pi}{a}\int_{BZ}\langle u_{k}|\frac{\partial}{\partial k}| u_{k}\rangle dk.
\label{eq:zakphase}
\end{equation}
 
According to Eq. (\ref{eqmoto2}), the quasimomentum center $k_{c}$ moves linearly across the first Brillouin zone,  $T_{B}=h/aF$ 
being the time needed to scan the full zone. The dynamics in real space is instead determined by the velocity term in the right side of  in Eq. (\ref{eqmoto1}), 
composed by the sum of the usual ``normal'' velocity
\begin{equation}
v_{N}(t)\equiv \frac{1}{\hbar}\frac{\partial\varepsilon(t)}{\partial k}\Big|_{k=k_{c}}
\label{velnormale}
\end{equation}
and of the \textit{anomalous velocity}
\begin{equation}
v_{A}(t)\equiv \frac{\partial A_{k}}{\partial t}\Big|_{k=k_{c}}-\frac{\partial\chi_{k}}{\partial k}\Big|_{k=k_{c}}.
\label{velanomala}
\end{equation}
The latter represents the Berry correction to the standard equation of motions \cite{ashcroft}, and it
appears as an electric field term in quasi-momentum
 space \cite{lorentz}.  
The anomalous velocity can be cast into the form 
\begin{equation}
v_{A}(t)=-\frac{4\pi}{a}{\rm Im}\langle \frac{\partial u_{k}}{\partial t}|\frac{\partial u_{k}}{\partial k}\rangle\Big|_{k=k_{c}}
\label{eq:va}
\end{equation}
that is manifestly vanishing for stationary potentials. In that case the wavepacket performs the usual Bloch oscillations, of period $T_{B}$ and amplitude 
$\Delta/F$, $\Delta$ being the energy bandwidth. Instead, in the case of time dependent parity-breaking  potentials, the effects of
 the anomalous velocity may become relevant, as we discuss in the following.
 
\section{The model}
\label{sec:model}

\begin{figure}[t]
\includegraphics[width=0.8\columnwidth]{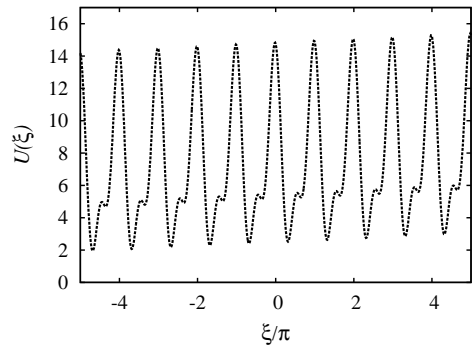}
\caption{Plot of the total potential $U(\xi)={\cal{V}}(\xi,0)+
{2\xi}/{\tau_{B}} $ at $\tau=0$, for ${\cal V}_{1}=10, {\cal V}_{2}=5, \theta=0.9\pi$.}

\label{fig:potenergy}
\end{figure}

In order to illustrate the effects of the anomalous velocity in a specific case, we consider the following periodic potential 
\begin{eqnarray}
V(x,t)&=&V_{1}(t)\cos^{2}\left(qx\right)+ V_{2}(t)\cos^{2}\left(2q x+\theta\right)
\label{eq:potential}
\end{eqnarray}
with period $a=\pi/q$. 
In the presence of an external force $F$, the hamiltonian takes the form
\begin{equation}
H=-\frac{\hbar^{2}}{2 m}\frac{\partial^2}{\partial x^2}+V(x,t) + Fx
\label{eq:ham}
\end{equation}
$m$ being the particle mass. Here we will consider the specific case of gravity, $F=mg$.
We also notice that the length $q^{-1}$, and the corresponding energy scale $E_{R}={\hbar}^2q^2/2 m $, represent two natural scales for the system, and it is therefore useful 
to rewrite Eqs. (\ref{eq:potential})-(\ref{eq:ham}) in a dimensionless form as
\begin{equation}
{\cal{H}}=-\frac{\partial^2}{\partial\xi^2}+{\cal{V}}(\xi,\tau)+\frac{2\xi}{\tau_{B}} 
\label{eq:dimham}
\end{equation}
with
\begin{equation}
{\cal{V}}(\xi,\tau)={\cal{V}}_{1}(\tau) \cos^{2}\left(\xi\right) + {\cal V}_{2}(\tau) \cos^{2}\left(2\xi+\theta\right)
\label{eq:pot}
\end{equation}
where  ${\cal V}_{i}=V_{i}/E_{R}$, $\xi=qx$, $\tau=E_{R} t/\hbar$, and $\tau_{B}=2\pi E_{R}/mga$ is the dimensionless Bloch period, which corresponds to the inverse of the normalized gravitational energy per unit cell, times $2\pi$. The typical shape of the potential is shown in
 Fig. \ref{fig:potenergy}. Here we are mainly concerned with the single band approximation, and in the
 following we will refer to the (dimensional)  bandwidth $\Delta$ of the lowest band as \textit{the
 bandwidth}, and to the (dimensional) minimal energy gap $E_{g}$ between the first and second band as
 \textit{the gap}.

\begin{figure}
\includegraphics[width=0.7\columnwidth]{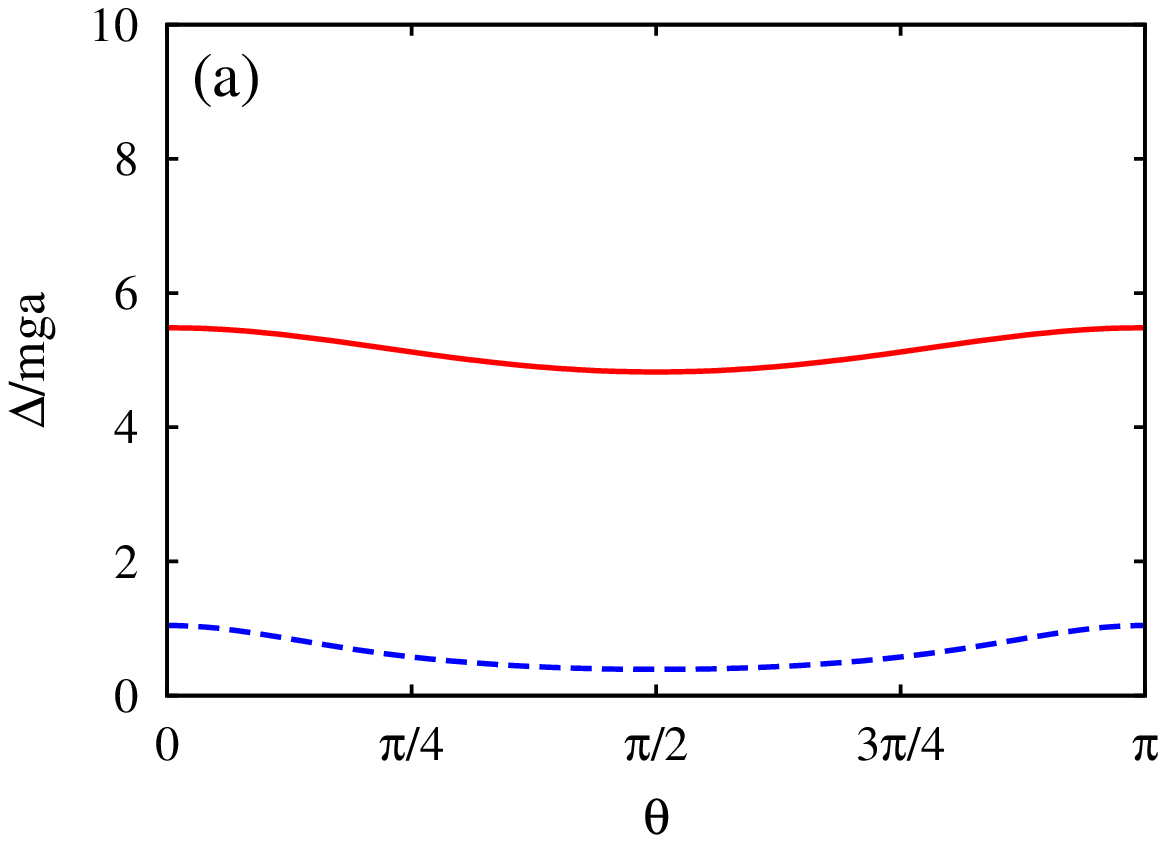}
\includegraphics[width=0.7\columnwidth]{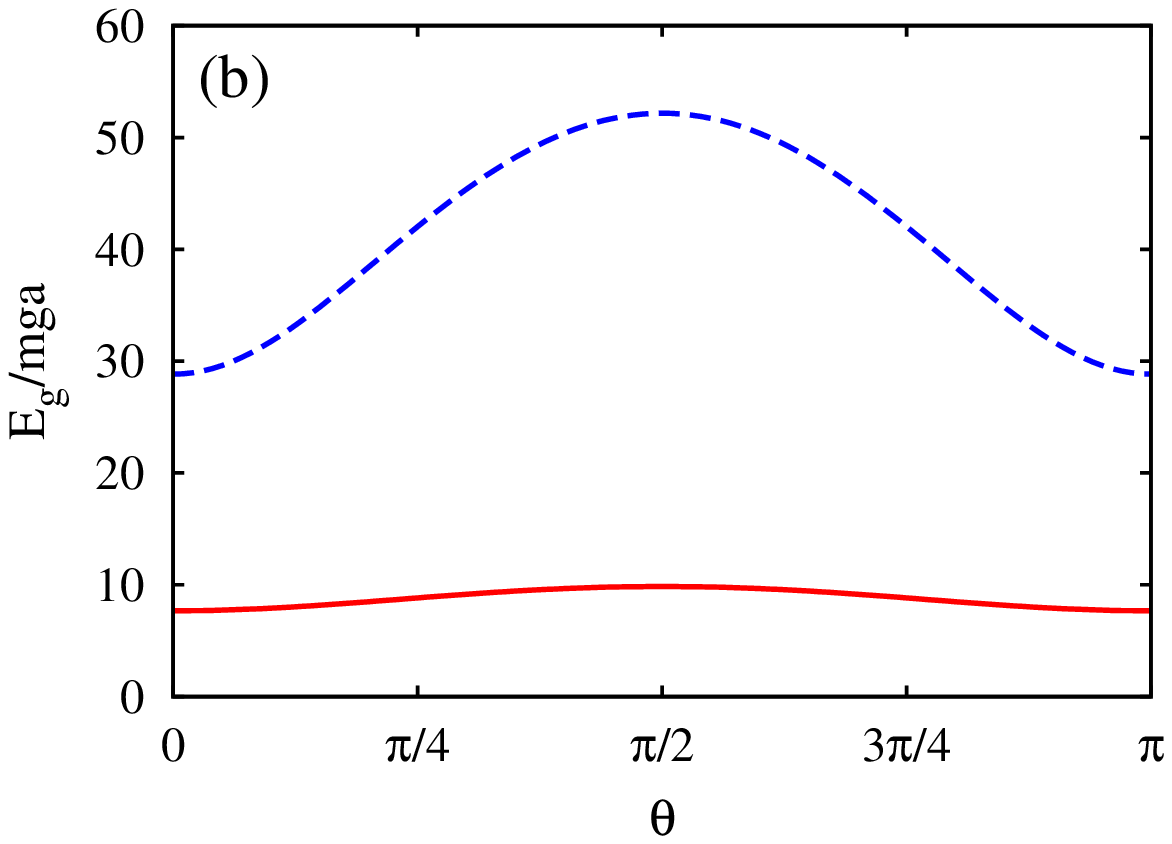}
\caption{(Color online) Bandwidth $\Delta$ (a) and minimal bandgap $E_{g}$ (b), normalized to the gravitational  energy per unit cell, as a function of $\theta$, for ${\cal V}_{1}={\cal V}_{2}=2$ (red, continuous line) and ${\cal V}_{1}=10,{\cal V}_{2}=5$ (blue, dashed line).}
\label{fig:band}
\end{figure}

As for the parameters, we choose typical values of the experiments with ultracold atoms in optical lattices \cite{morsch}. We fix $\tau_{B}/2\pi=10$. Then, ${\cal{V}}_{1}$ is chosen in the range $2\div10$ in order to have - at least in the 
case of a static primary lattice alone - clean Bloch oscillations (no drifts), large band gap (in order to conform to the single band approximation), 
and large enough oscillation amplitudes (i.e. large enough bandwidths $\Delta$).
The latter condition is obtained by considering ${\cal V}_{1}\leq 10$.
Conversely, the lower bound ${\cal{V}}_{1}=2$ guarantees that the ratio of the recoil energy to the gravitational energy is above the critical value for having no drift,
  $\tau_{B}{\cal{V}}_{1}/2\pi\approx20$. In addition, for this value the minimal gap $E_{g}$ is already one
 order of magnitude greater than  the gravitational energy per site $mga$ (see Fig. \ref{fig:band}), and
 this accounts for the use of the single band approximation (actually, this corresponds to the
 adiabatic approximation, $T_{B}>h/E_{g}$). 

Let us now consider the effect of the secondary lattice. For ${\cal V}_{2}\neq0$, both the bandwidth $\Delta$ and the 
gap $E_{g}$ strongly depend on $\theta$, as shown in Fig. \ref{fig:band} \cite{bloch_numerics}. The band structure is 
periodic with period $\pi$, and is characterized by parity centers at integer multiples of $\pi/2$.
In addition, $\Delta$ ($E_{g}$) also depend on the secondary lattice intensity, increasing (decreasing) monotonically as ${\cal V}_{2}$ is increased (in the range considered here). In the following we consider the case ${\cal V}_{2}<{\cal V}_{1}$.

\begin{figure}[t!]
\includegraphics[width=0.8\columnwidth]{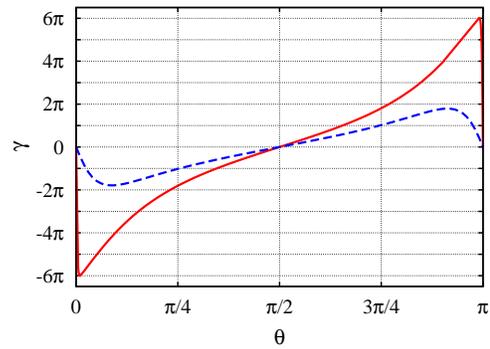}
\caption{(Color online) Zak-Berry phase $\gamma$ as a function of $\theta$, for ${\cal V}_{1}={\cal V}_{2}=2$ (red, continuous line) and
${\cal V}_{1}=10,{\cal V}_{2}=5$ (blue, dashed line), as obtained from from Eq. (\ref{eq:zakphase}).}
\label{fig:zak}
\end{figure}

For $\theta\neq n\pi/2$ ($n$ integer) parity is broken, and this is signaled by the appearance of a non vanishing
Zak-Berry phase $\gamma$, defined in Eq. (\ref{eq:zakphase}). This quantity is proportional to the offset of the Wannier
 function centers \cite{libro,marzari} 
\begin{equation}
\gamma=\frac{2\pi}{a}\bra{w_{0}}x\ket{w_{0}} 
\label{zakeq}
\end{equation}
($w_{0}$ being the Wannier functions located at site $j=0$ \cite{callaway}) modulo a vector of the direct lattice, coming from equivalent choices of the Bloch basis
(with $e^{\displaystyle{i\gamma}}$ being the gauge invariant quantity).
The Zak-Berry phase $\gamma$ (calculated from Eq. (\ref{eq:zakphase})) is shown in Fig. \ref{fig:zak}, as a function of $\theta$,
for two different sets of the potential intensities. 
In both cases, $\gamma$ grows monotonically reaching its maximum at  $\theta\approx0.9\div0.95\pi$, and then goes steeply to zero at  $\theta=\pi$.  The same behavior can be found in a wide range of values of ${\cal V}_{i}$ ($i=1,2$). This suggests that the maximal effects of the anomalous terms in the dynamics should be expected for these values of $\theta$, as we will discuss in the next section.
In addition, we find that the Zak-Berry phase is almost independent of ${\cal V}_{1}$, and slightly decreases as ${\cal V}_{2}$ is increased.

This analysis reveals that the control of the phase $\theta$ is rather a crucial issue for commensurate bichromatic lattices, since the band 
properties have a strong dependence on it.
This may become even more relevant in higher dimensions, where parity breaking alone (no need for time-dependent bands) is sufficient for producing
anomalous terms in the semiclassical equation of motions or other effects related to the Berry phase \cite{niu}.

\section{Anomalous Bloch oscillations}
\label{sec:bloch}

Let us now turn to the dynamical behavior of the system, by considering the evolution of an initial wavepacket under the effect of a constant force, and in the presence of a modulation of the periodic potential. 
In particular, we consider a gaussian wavepacket modulated by the periodic potential, obtained as the ground-state of the potential ${\cal V}(\xi,0)$ plus an additional harmonic confinement ${\cal V}_{ho}(\xi)=\alpha \xi^{2}$ (we chose $\alpha=3\cdot 10^{-5}$, that corresponds to having about twelve occupied lattice sites) \cite{numeric2}.
Then, the wavepacket evolution is investigated by solving the Schr\"odinger equation \cite{numeric1}
\begin{equation}
i\partial_{\tau}\psi(\xi,\tau)={\cal{H}}(\tau)\psi(\xi,\tau)
\label{eq:schrod}
\end{equation}
with ${\cal{H}}(\tau)$ in Eq. (\ref{eq:ham}).

As anticipated, the combined effect of parity breaking and time modulation gives rise to an anomalous  velocity term in the semiclassical 
equation for the center of mass dynamics in real space, see Eq. (\ref{eqmoto1}). The importance of this term can be analyzed by comparing the evolution of 
$\xi_{c}(\tau)\equiv\langle\psi|\xi|\psi\rangle_{\tau}$ obtained from the full solution of the Schr\"odinger equation, with that predicted 
by the semiclassical equations (\ref{eqmoto1})-(\ref{eqmoto2}) \cite{bloch_numerics}.

In particular, let us separate the trajectory of the wave packet center into two parts
\begin{equation}
\xi_{c}(\tau)=\xi^{N}_{c}(\tau)+\xi^{A}_{c}(\tau)
\label{center} 
\end{equation}
the first coming from the time integration of the ``normal'' velocity in eq. (\ref{velnormale}) and the second from the ``anomalous'' 
velocity in eq. (\ref{velanomala}). Then, we are interested in studying the role of the contribution $\xi^{A}_{c}(\tau)$ on the full solution
when the shape of the potential well is modulated in time, and to find
regions in the parameter space where this term strongly characterizes the dynamics.
We remind that Eq. (\ref{eqmoto2}) is not affected by anomalous terms, and the evolution
of the wavepacket center in quasimomentum space is linear in time; in dimensionless units it reads
\begin{equation}
\tilde{k}_{c}(\tau)= \tilde{k}_{c}(0) + 2\frac{\tau}{\tau_{B}}.
\label{eq:ktilde}
\end{equation}

Thus, we consider the following periodic modulations of the 
potential amplitudes
\begin{equation}
{\cal V}_{i}(\tau)={\cal V}_{i}\left(1+A_{i}  {\rm sin}^{2}\left(\Omega_{i}\tau\right)\right)
\label{timepot}
\end{equation}
where $A_{i}$ and $\Omega_{i}\equiv\eta_{i}\pi/\tau_{B}$ are the modulation amplitudes and frequencies, respectively.

\begin{figure}
\includegraphics[width=0.9\columnwidth]{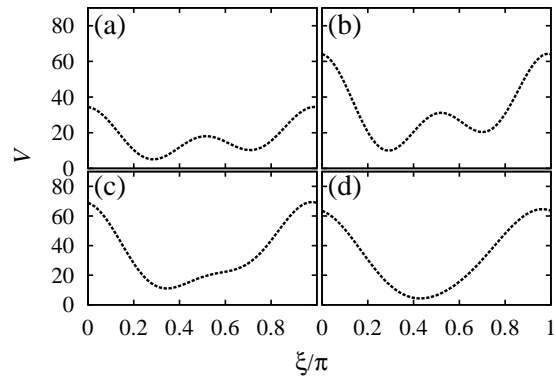}
\caption{Plot of the periodic potential ${\cal V}(\xi)$ at 
$\tau/\tau_{B}=1/8$ (a), $1/4$ (b), $3/8$ (c), $1/2$ (d), for 
${\cal V}_{1}=10$, ${\cal V}_{2}=5$, $\theta=0.9\pi$, $A_{1}=A_{2}=5$, $\eta_{1}=1$, $\eta_{2}=2$.}
\label{fig:frames}
\end{figure}

Here we consider rational values of $\eta_{i}$, corresponding to periodic modulations with periodicity multiple of the Bloch period, in order to have a cyclic motion in parameter (quasimomentum) space, as required by the Berry's theory. The upper bound on $\eta_{i}$, as well as that on the modulation amplitudes $A_{i}$, is given by the
 limit of validity of the one band approximation, 
which requires that the band gap during the dynamics evaluated at $k_{c}(t)$, $E_{g}^{c}(t)$,  remains much greater than the modulation energies.
This condition turns out to be satisfied, in a large region of the parameter space, also for values of $A_{i},\eta_{i}\gg 1$ \cite{nota_adiabatic}.
Typical shapes of the potential during one modulation period are shown in Fig. \ref{fig:frames}.

\begin{figure}
\includegraphics[width=0.7\columnwidth]{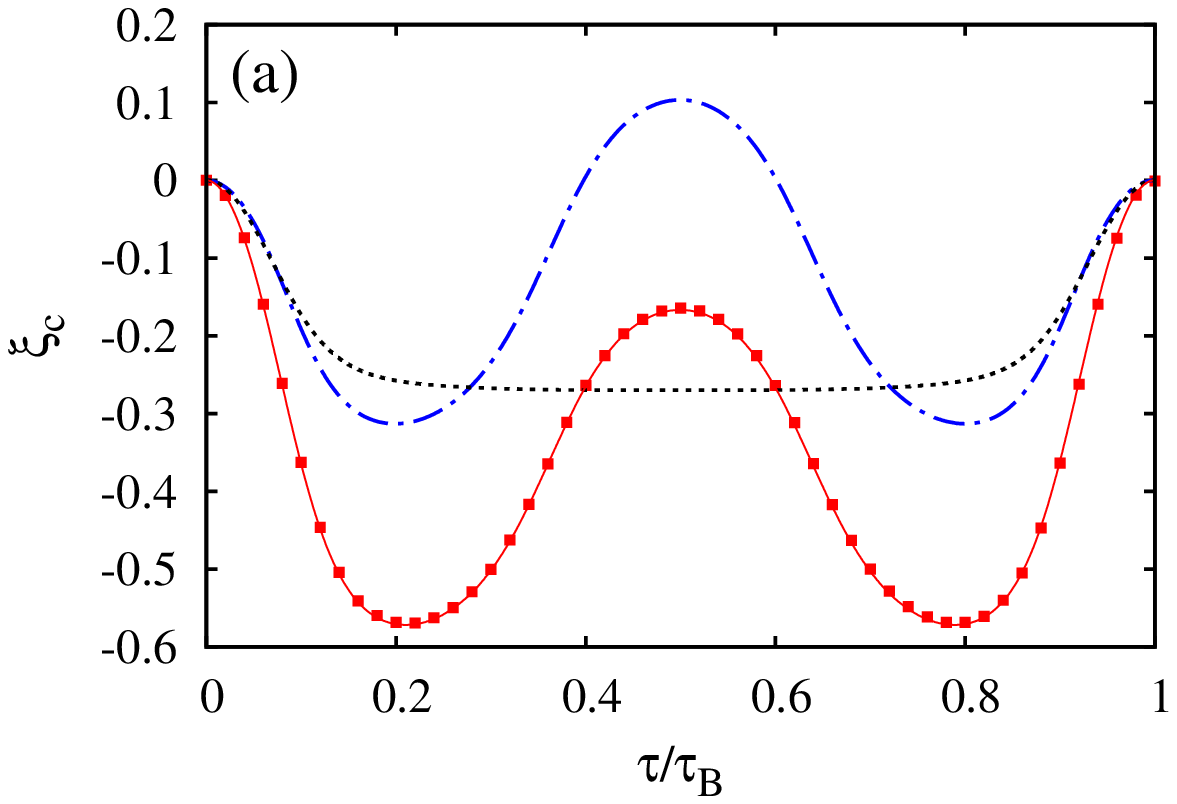}
\includegraphics[width=0.7\columnwidth]{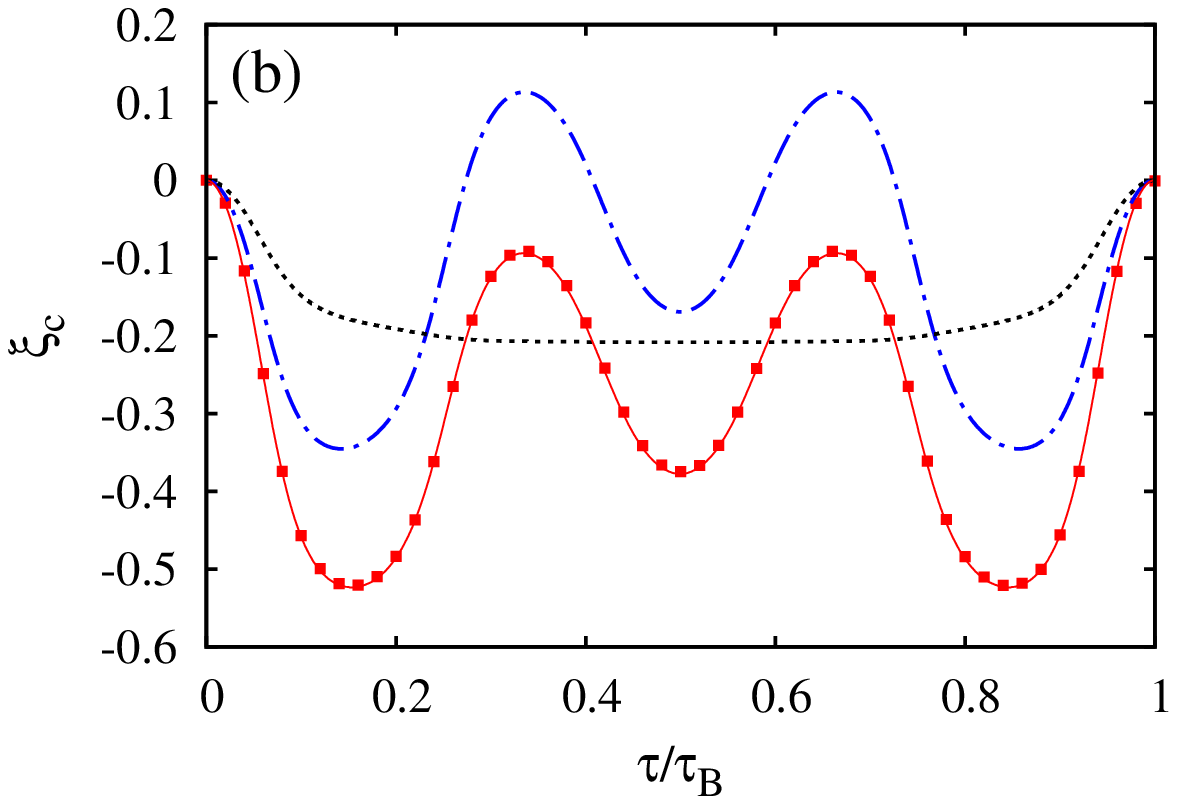}
\includegraphics[width=0.7\columnwidth]{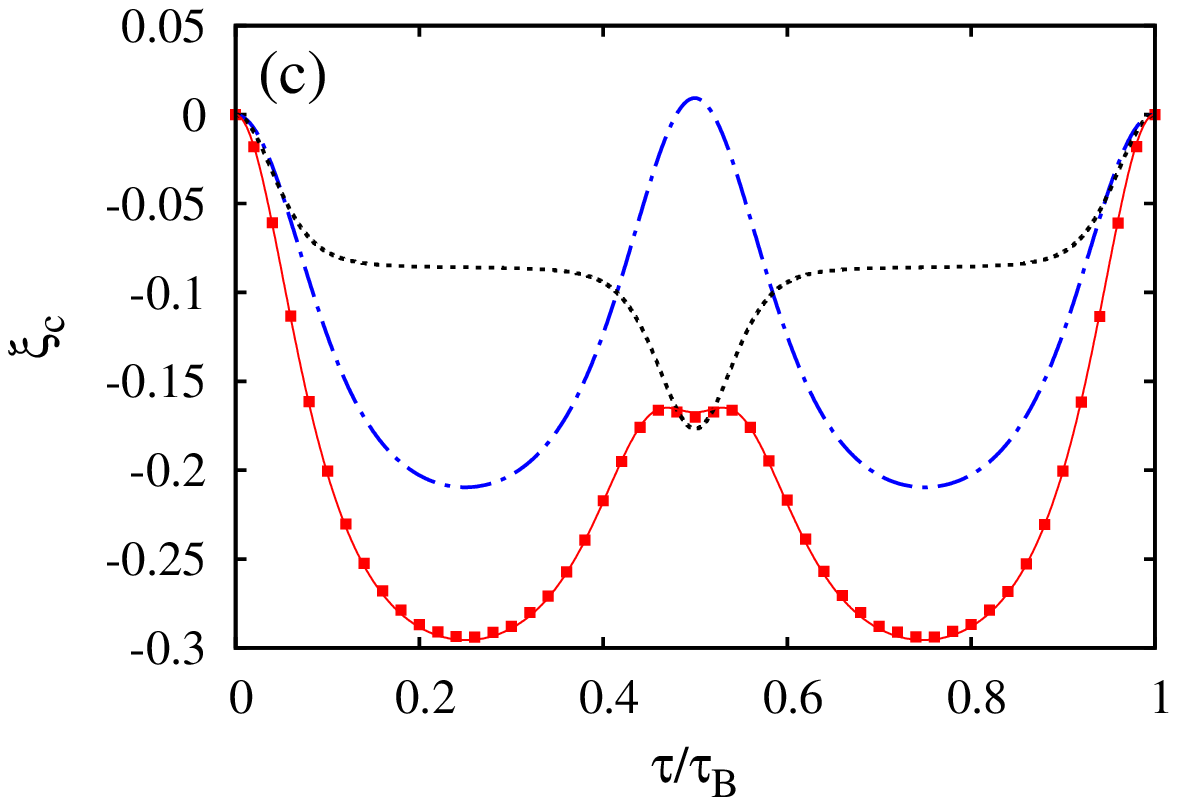}
\caption{(Color online) Evolution of the wave packet center $\xi_{c}$, for 
 $A_{1}=A_{2}=5$, $\eta_1=1$, $\eta_2=2$ (a), $\eta_{1}=1$,
 $\eta_2=3$ (b), and $\eta_1=\eta_{2}=2$ (c). The full semiclassical 
evolution (red, continuos line) perfectly fits the solution of the Schr\"odinger equation (red points). The terms corresponding to the normal dynamics, $\xi_c^{N}$ (black, dotted line), and anomalous dynamics, $\xi_c^{A}$ (blue, dashed line), are also shown.}
\label{fig:xc}
\end{figure}

We start by describing some specific examples, shown in Fig. \ref{fig:xc}. 
Unless explicitly stated, all the results presented in this section are obtained for the following set parameters characterizing the ``static'' properties of the potential:  
${\cal V}_{1}=10, {\cal V}_{2}=5, \theta=0.9\pi$. Then, we will discuss the dependence of the results on the choice of parameters, as well as their generality.

In Fig. \ref{fig:xc} we plot the evolution of the wave packet center $\xi_{c}(\tau)$, 
along with the separate contributions coming from the normal and anomalous terms,
 for 
$A_{1}=A_{2}=5$, $\eta_{1}=1$, $\eta_{2}=2$ (a), $\eta_{1}=1$, $\eta_{2}=3$ (b), and $\eta_1=\eta_{2}=2$ (c). The full semiclassical evolution of Eq. (\ref{eqmoto1}) (red, continuos line) perfectly agrees with the solution of the Schr\"odinger equation (red points). For $\eta_{1}\neq\eta_{2}$, the normal term $\xi_{c}^{N}(\tau)$ (black, dotted line) shows peaks only at $\tau=0$ and $\tau=\tau_{B}$ whereas the anomalous term $\xi_{c}^{A}(\tau)$ (blue, dashed line) shows one (a) or two (b) additional central peaks. In general, we find that the number of central peaks (excluding  those at $\tau=0$ and $\tau=\tau_{B}$) equals $\eta_{2}-1$. This holds also for higher integer values of $\eta_2$, provided that the single band approximation is satisfied.
The dependence on $\eta_{1}$ at fixed $\eta_{2}$ is more weak,
affecting the trajectory without basically changing its structure. 
In particular, when $\eta_{1}=\eta_{2}$ as in Fig. \ref{fig:xc}c, the modulation does not change the shape of the potential (if $A_{1}=A_{2}$). In this case we find an additional central peak also in the normal term, in the opposite direction with respect to the anomalous component. 

We notice that in all these examples the anomalous term, $\xi_{c}^{A}(\tau)$, dominates over the normal one, $\xi_{c}^{N}(\tau)$. In general, the relative weight between the two depends on the amplitudes $A_{i}$, as will be discussed later on.

We have also analyzed cases in which $\eta_{1},\eta_{2}\le 1$, finding that the qualitative picture does not change. The significant 
difference is that the periodicity of the motion amounts to several Bloch periods, as expected. In addition, even larger values of $A_{i}$ can be choosen without breaking the single band approximation.

\begin{figure}
\includegraphics[width=0.7\columnwidth]{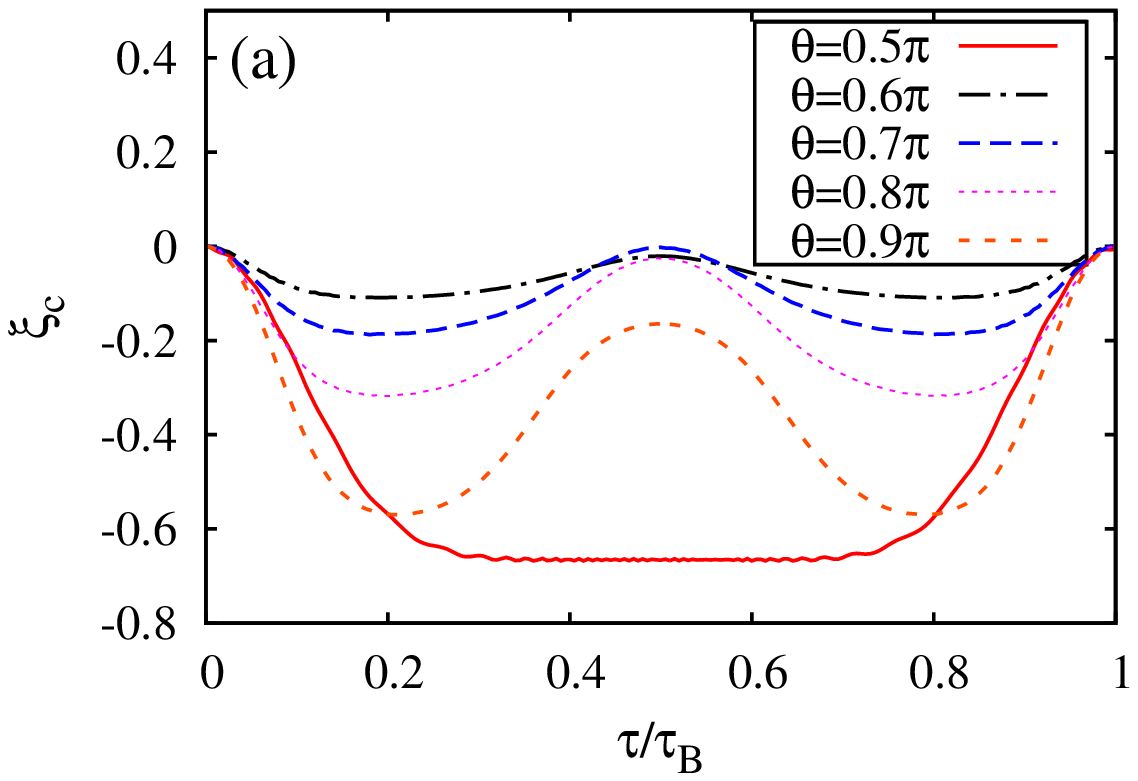}
\includegraphics[width=0.7\columnwidth]{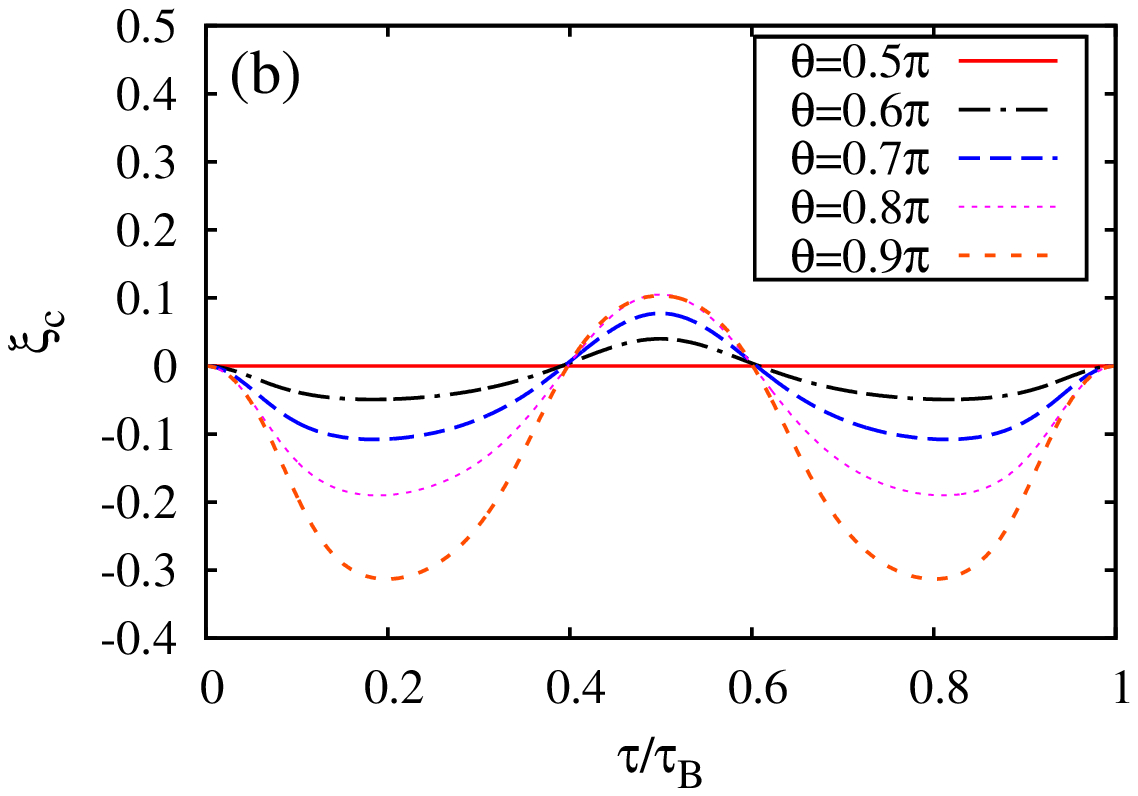}
\caption{(Color online) (a) Evolution of the wave packet center $\xi_{c}(\tau)$ for $A_{1}=A_{2}=5$, $\eta_1=1,\eta_{2}=2$ and different values of $\theta$. (b) The contribution of the anomalous term $\xi_{c}^{A}(\tau)$.}
\label{fig:diptheta}
\end{figure}

An interesting issue concerns the dependence of these dynamical results on $\theta$, which can be considered as a dependence on the degree of  parity breaking. In Fig. \ref{fig:diptheta} we plot the evolution of the wave packet center $\xi_{c}(\tau)$ by varying $\theta$, for  $A_{1}=A_{2}=5$, $\eta_1=1$ and $\eta_{2}=2$. In panel (a) we
 report the full solution $\xi_{c}(\tau)$, while in (b) just the contribution of the anomalous term $\xi_{c}^{A}(\tau)$. The values of
 $\theta$ correspond to a growing Zak-Berry phase (see Fig. \ref{fig:zak}) from $\gamma=0$ (no parity breaking) at $\theta=0.5\pi$, towards its maximum value around $\theta=0.9\pi$. It is evident that, as anticipated in the previous section, the amplitude of the anomalous oscillations grows together with $\theta$ in this interval, showing the relevance on the degree of parity breaking on the resulting dynamics. It is worth mentioning that in the interval of negative Zak-Berry phases, that is for $\theta\in (0,\pi/2)$ the behaviour is almost specular. Indeed, the two cases are connected by a mirror transformation  $\theta\rightarrow \pi-\theta$; though this is not an exact symmetry due to the presence of gravity, the qualitative picture does not change in the two cases.

\begin{figure}
\includegraphics[width=0.7\columnwidth]{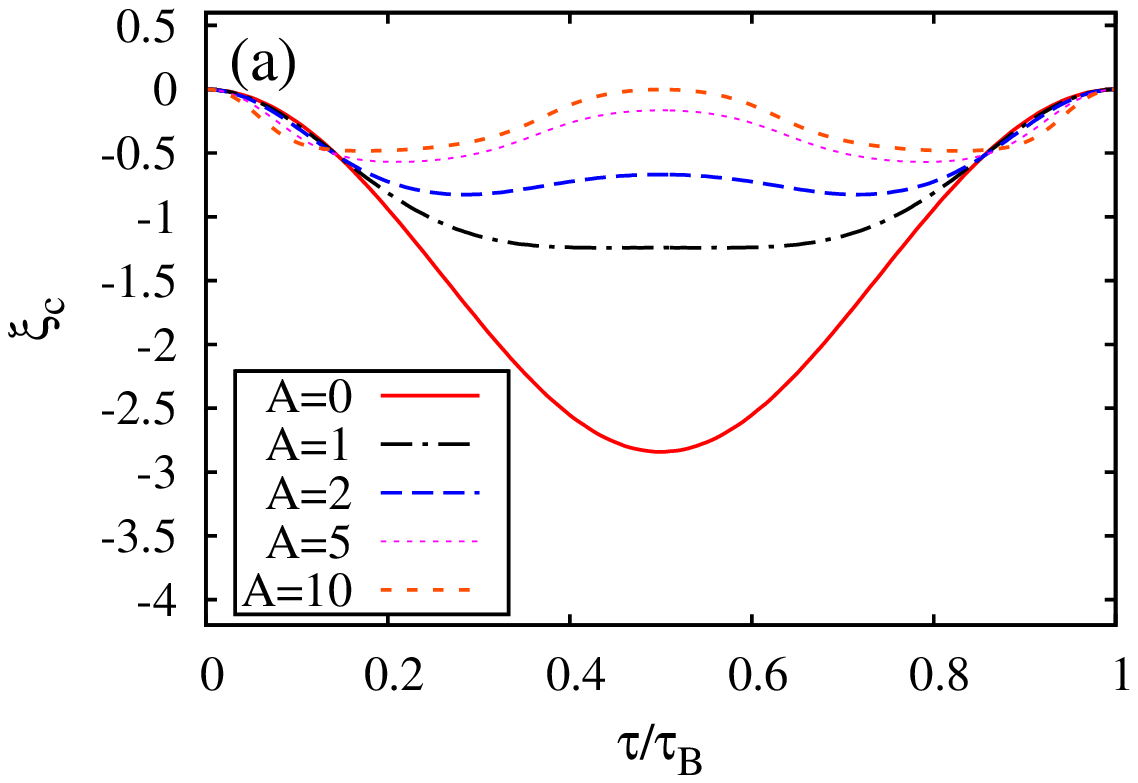}
\includegraphics[width=0.7\columnwidth]{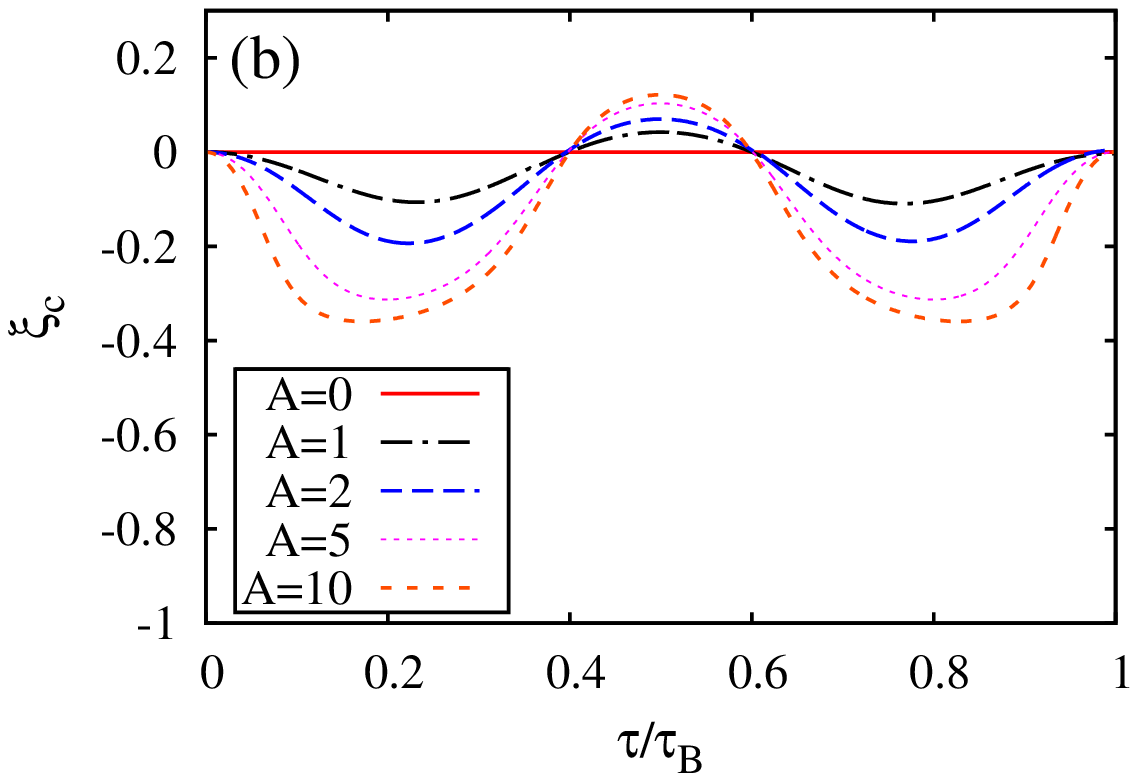}
\caption{(Color online) (a) Evolution of the wave packet center $\xi_{c}(\tau)$ for $\eta_1=1,\eta_{2}=2$ and different values of  $A=A_{1}=A_{2}$. (b) The contribution of the anomalous term $\xi_{c}^{A}(\tau)$.}
\label{fig:dipa}
\end{figure}

Let us now consider the dependence on the modulation amplitudes $A_{i}$. Since the 
anomalous velocity contains time derivatives of the periodic Bloch wave functions, we expect the anomalous effects to increase as $A_{i}$ is increased. 
This is indeed the case, as shown in Fig. \ref{fig:dipa} where we plot the full solution $\xi_{c}(\tau)$ (a) and just the anomalous term   $\xi_{c}^{A}(\tau)$ (b), for different values of $A=A_{1}=A_{2}$, and  $\eta_{1}=1$, $\eta_{2}=2$. This figure shows that, for $A$ in the range $[0,10]$, the amplitude of anomalous oscillations grows monotonically as expected (we find that in general higher values of $A$ are beyond the validity of the single band approximation). In addition, we find that while the amplitudes of normal Bloch oscillations (for $A=0$) can be of the order of some lattice site, those of anomalous oscillations are rather smaller. In particular, 
anomalous oscillations become predominant only when the bandwidth is small (that is, for rather large ${\cal {V}}_{1}(t)$).

Finally, we have also studied the case $A_{1}\neq A_{2}$, finding a more pronounced dependence of the results on $A_{2}$ rather than on $A_{1}$, according to what previously found for $\eta_{1}$ and $\eta_{2}$. This is not surprising, since parity is broken due to
 the presence of secondary lattice, and consequently its parameters mainly influence the anomalous
dynamics.

\begin{figure}
\includegraphics[width=0.8\columnwidth]{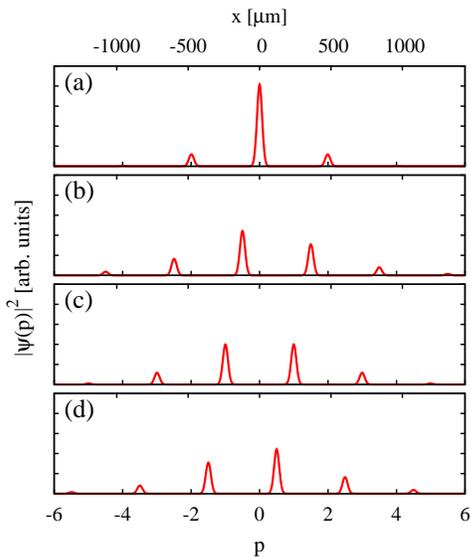}
\caption{(Color online) Plot of the momentum distribution $|\Psi(p)|^{2}$ at 
$\tau/\tau_{B}=0$ (a), $1/4$ (b), $1/2$ (c), $3/4$ (d), for the case shown in Fig. \ref{fig:xc}a.
The horizontal axis on the top represents the distance after the time-of-flight (see text).}
\label{fig:ft}
\end{figure}

\section{Measuring the anomalous velocity}
\label{sec:av}

\begin{figure}
\includegraphics[width=0.99\columnwidth]{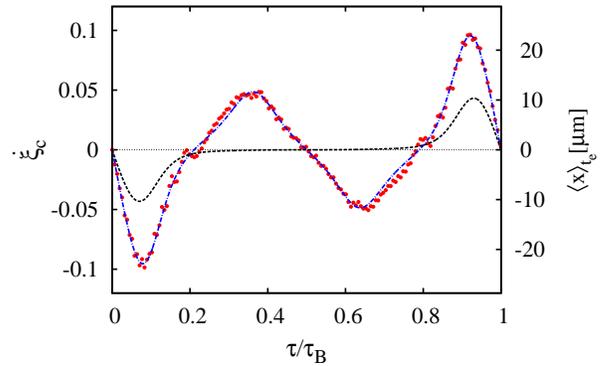}
\caption{(Color online) The wavepacket velocity for the case shown in Fig. \ref{fig:xc}a. ($A_{1}=A_{2}=5$, $\eta_1=1$, $\eta_2=2$).
The semiclassical velocity $\dot{\xi}_c$ (red, continuos line) fits that extracted from the momentum distribution of the full Schr\"odinger solution (red points). The normal velocity $\dot{\xi}_c^{N}$ is also shown (black, dashed line). The right vertical scale corresponds to the center $\langle x\rangle_{t_{exp}}$ of the density distribution after a time of flight $t_{exp}=50$ ms for a non-interacting condensate of $^{39}$K.}
\label{fig:av}
\end{figure}
Let us now discuss the experimental relevance of these results, focusing on current experiments with
ultracold atoms in optical lattices. As we have previously discussed, the amplitude of anomalous Bloch oscillations in coordinate space is rather small (only a fraction of a lattice size), and   therefore a direct detection by \textit{in-situ} imaging is not feasible, due to resolution limitations.
Usually, Bloch oscillations are detected via time-of-flight measurements, probing momentum space. Typical shapes of the momentum distribution $|\Psi(\tilde{p})|^{2}$ are shown in Fig. \ref{fig:ft} (for the wavepaket in Fig. \ref{fig:xc}a) . It is characterized by sharp peaks at $\tilde{p}(\tau)=\pm2n +\tilde{k}_{c}(\tau)$ ($n=0,\pm1,\dots$) \cite{pedri,nesi}, with $\tilde{k}_{c}(\tau)$ being the wavepaket quasimomentum given by Eq. (\ref{eq:ktilde}). This equation is not affected by anomalous terms, and therefore it is not possible to measure the effect of the Berry terms simply from the shift of the peaks, as they move linearly in time as for normal Bloch oscillations.
Nevertheless, their relative population depends on the modulation, and the wavepacket velocity $\dot{\xi}_{c}(\tau)$ can be extracted from the momentum distribution, according to Ehrenfest theorem $d\langle{x}\rangle/dt=\langle{p}\rangle/m$, that in dimensionles units reads
\begin{equation}
\dot{\xi}_{c}(\tau)=\frac12 \langle{\tilde{p}}\rangle(\tau).
\end{equation}

In the experiments, the momentum distribution is inferred from the density distribution
after the time of flight, owing to the relation
\begin{equation}
|\psi_{\infty}(x,t)|^{2}\equiv\lim_{t_{e}\to\infty}|\psi(x,t;t_{e})|^{2}\propto|\Psi(mx/t_{e},t)|^{2}
\end{equation}
where $t_{e}$ is the free expansion time, and $t$ the evolution time in the trap.
Therefore, the wavepacket velocity $\dot{\xi}_{c}(\tau)$ after an oscillation time $\tau=E_{R} t/\hbar$ can be measured from the center $\langle x\rangle_{t_{e}}\equiv\langle\psi(t;t_{e})|x|\psi(t;t_{e})\rangle$ of the measured density distribution after a time of flight $t_{e}$, as 
\begin{equation}
\dot{\xi}_{c}(\tau)=\frac{2}{\hbar q}\frac{m}{t_{e}}\langle x\rangle_{t_{e}.}
\end{equation}

In Fig. \ref{fig:av} we show the comparison of the semiclassical velocity $\dot{\xi}_{c}(\tau)$
with that extracted from the momentum distribution, $0.5\langle{\tilde{p}}\rangle(\tau)$, and the corresponding center $\langle x\rangle_{t_{e}}$ of the
 density distribution after a time of flight $t_{e}=50$ ms, for a non-interacting $^{39}$K condensate \cite{roati}. In this case, $\langle x\rangle_{t_{e}}$ 
turns out to be of the order of some tens of microns, and its detection is experimentally feasible. In addition, the effects of the anomalous velocity 
term are pronounced in a large time range where the normal velocity is almost vanishing, and this should help their visibility. 

\begin{figure}
\includegraphics[width=0.9\columnwidth]{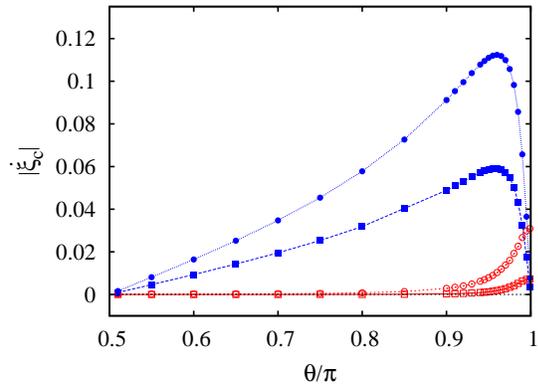}
\caption{(Color online) Plot of maximum of the modulus of the normal and anomalous velocities, $|\dot{\xi}_c^{N}|$ (empty symbols)  and $|\dot{\xi}_c^{A}|$ (filled symbols) respectively, as a function of $\theta$, for $\eta_1=1$, $\eta_2=2$ (squares) and $\eta_1=1$, $\eta_2=3$ (circles). In all cases $A_{1}=A_{2}=5$.}
\label{fig:av-theta}
\end{figure}
As discussed before, these effects can be maximized by chosing suitable values of the angle $\theta$.
 In Fig. \ref{fig:av-theta} we plot the maximum of the anomalous and normal velocities ($\dot{\xi}_c^{A}$ and $\dot{\xi}_c^{N}$, respectively) 
 as a function of $\theta$, for $0.5\pi<\theta<\pi$ and two different sets of $\eta_{i}$'s. In particular,
 it is convenient to choose $\theta\approx0.9\pi$, where the magnitude of $\dot{\xi}_c^{A}$ is close to its maximun and $\dot{\xi}_c^{N}$ almost vanishing.

\section{Conclusions}
\label{sec:conclusions}

We have investigated the occurrence of anomalous Bloch oscillations in the dynamics of a wave packet in a parity-breaking one-dimensional periodic potential that is modulated in time, in the presence of an additional weak linear potential. We have considered a periodic potential composed by the sum of two sinusoidal potentials of commensurate wavelengths, shifted by an angle $\theta$ that characterizes the parity breaking.
We have found that the anomalous velocity correction of the semiclassical equations - related to Berry's phase - profoundly affects the wave packet dynamics, in agreement with the full solution of the Schr\"odinger equation. We have investigated how these effects depend on $\theta$, finding that the maximal effects are for $\theta\approx0.9\pi$ both for static and dynamical properties.
These results are an explicit demonstration of the importance of Berry's corrections beyond the usual semiclassical approximation. In particular, these effects may become relevant for current experiments with ultracold atoms in optical lattices, when also time modulations and parity breaking are considered. As an example, we have shown that the effects of the anomalous velocity can be measured from the atomic distribution after the time-of-flight, in current experiments with non-interacting Bose-Einstein condensates in bichromatic optical lattices.

\begin{acknowledgments} 
We thank R. Franzosi for fruitful discussions, and G. Modugno for useful suggestions. 
\end{acknowledgments}


\begin{thebibliography}{}
\bibitem{bloch_numerics}
The spectrum and the anomalous velocity term are computed by using the truncated expression
$u_{k}(\xi)=\sum_{\ell=-r}^{r}\frac{1}{\sqrt{2\pi}} \psi_{\ell}(k) e^{\displaystyle{i{\ell \xi}}}$, with $r\gg 1$,  and then solving the resulting eigenvalue 
equations for the coefficients $\psi_{\ell}$ by means of LAPACK routines.
\bibitem{nota_adiabatic}
From our simulations we have found that the adiabatic approximation works well beyond then naive expectation.
\end{thebibliography}
\end{document}